\documentstyle[aps]{revtex}

\input{epsf}

\begin{document}
\title{{TOPOLOGICAL CORRELATIONS IN SOAP FROTHS}}   
\author{{K.Y. Szeto, T. Aste* and W. Y. Tam} }
\address{Department of Physics, The Hong Kong University of Science 
and Technology, phszeto@usthk.ust.hk} 
\address{Clear Water Bay, Kowloon, Hong Kong.}
\address{*Equipe de Physique Statistique, LUDFC  Universit\'e Louis Pasteur,}
\address{ 3, rue de l'Univresit\'e, 67084 Strasbourg 
France. tomaso@ldfc.u-strasbg.fr}

\date{\today}
\maketitle

\begin{abstract}
Correlation in two-dimensional soap froth is analysed with 
an effective potential for the first time. 
Cells with equal number of sides repel (with linear correlation) 
while cells with different number of sides attract (with NON-bilinear) 
for nearest neighbours, 
which cannot be explained by the maximum entropy argument. 
Also, the analysis indicates that froth 
is correlated up to the third shell neighbours at least, 
contradicting the conventional ideas that froth is not strongly correlated. 
\end{abstract}


\bigskip

Two-dimensional cellular structures are space-filling disordered partitions 
of space by cells that are irregular polygons 
and often appear in nature as in metal 
grains, biological tissues and common soap froths \cite{Stavans:0}.  
These cellular patterns are usually trivalent 
($3$ edges meet at a vertex) due to topological stability. 
A generic term for these networks is ``froth'', since 
the soap-froth is the archetype of such structures.
Many theoretical studies on froth are based on 
mean field, with minimum consideration on the correlation 
between cells \cite{Stavans:4,Flyvberg}. 
This is understandable as the conventional wisdom assumes 
small correlation effect in froth, which is 
almost completely described by the Aboav-Weaire law \cite{Aboav,Wea,GKY}.  
Also, any theoretical description taken correlation effects into account 
is much more difficult. 
Thus, most of the approaches on froths 
make some kind of independent bubble approximation, and 
neighbours are described by a mean field\cite{Stavans:4} 
or as random neighbours\cite{Flyvberg}. 
It is our aim to test these approaches 
by analysing the experimental data on real dry soap 
froth and by extracting the topological correlation function between cells. 
We find that there are very strong and long range correlation, 
implying independent bubbles approaches are unrealistic.
The data indicate that any reasonable description respecting the correlation 
effects on froth must go at least to the third shell neighbours. 
Furthermore, these correlations effects are not at all 
well described by the maximum entropy arguments \cite{rivier}, 
as the predicted bilinearity for the correlators 
is only a rather crude approximation to the data. 
The data therefore strongly suggest that there must be a 
new theory describing real soap froth, with full consideration 
of the energies and the correlations. 

In froths a cell can be characterized topologically by its number of sides. 
Froths are disordered systems and a topological, statistical description 
of their structure is given by the correlation function. 
To extract the relevant structural properties 
from a disordered system like a froth is a very difficult task.
Indeed, the absence of symmetries and periodicities requires a priori 
the knowledge of the information about all the 
cells in the system.
A simple and useful approach is to analyze the 
froth as structured in concentric layers of cells at 
the same topological distance $j$ from a given central cell 
\cite{Forte,szetotam1,AST,AsBoRi} (where the topological distance 
between two cells is the minimum number of edges that a path must 
cross to go from one cell to the other \cite{szetotam1,AST,AsBoRi}).
An interesting and meaningful quantity in this analysis 
by concentric layers is, for instance, the number of 
cells in a given layer ($K_j(n)$, with $n$ the 
number of sides of the central cell)\cite{AST}.
Here we focus on the correlation function. 
The two-cell topological correlation function ($C_j(n,m)$) is 
the probability of finding a cell with $m$ sides at a distance $j$ 
from a cell with  $n$ sides.
The two-cell topological correlation function is 
equal to the total number  $N_j(n,m)$   of  
couples of cells with respectively $n$ and $m$ sides which 
are at a relative distance $j$, divided by the number of all couples 
of cells at a relative distance $j$ 
\begin{equation}
C_j(n,m)={N_j(n,m) \over 
\sum_{n,m=3}^\infty N_j(n,m)  } \;\;\;\;.
\label{Cnm}
\end{equation}
We consider the $n-$ and $m-$ sided cells as distinguishable.
Therefore each couple is counted twice and $N_j(n,m)$ 
and $C_j(n,m)$ are symmetric 
in $n, m$ (i.e. $C_j(n,m)=C_j(m,n)$).

By definition, in uncorrelated systems the 
correlation function must factorize: $C_j(n,m) = s_j(n)s_j(m)$.
The normalization and the symmetry of $C_j(n,m)$ imply $\sum_m s_j(m) =1$, 
consequently $s_j(n) = \sum_m C_j(n,m)$.
Now, we calculate this quantity and we interpret it in term of the properties 
of the layers.
Consider two cells with $n$ and $m$ sides 
which are at a relative topological distance $j$. 
Such a pair, connected by a path of length $j$, can be seen as a ``string'' 
with $n$- and $m$-sided terminations. 
Clearly the number of these strings in the cellular system 
is $N_j(n,m)$ (where each string is counted twice 
since the terminations are distinguishable cells). 
The number of strings 
of length $j$ with a termination in a cell with  
$n$ sides and the other termination free is given by  $\sum_m N_j(n,m)$. 
The same quantity is equal to the number of $n$-sided 
cells in the system ($N(n)$) multiplied by the average 
number of strings of length $j$ which terminate in one 
$n$-sided cell ($K_j(n)$). 
Therefore, $\sum_{m=3}^\infty N_j(n,m) = N(n) K_j(n).$
Finally, the total number of strings of length $j$ 
is equal to the total number of 
cells ($N_T$) multiplied by the average 
number of strings of length $j$ which terminate in any given cell 
($\langle K_j \rangle$), giving 
\begin {equation}
\sum_{n,m=3}^\infty N_j(n,m) 
= \sum_{n=3}^\infty N(n)K_j(n) 
= N_T \sum_{n=3}^\infty {N(n) \over N_T} K_j(n) 
= N_T \langle K_j \rangle  \;\;\;\;,
\label{N2}
\end{equation}
where $\langle (...) \rangle = \sum_n p(n) (...)$, with $p(n)=N(n)/N_T$ 
being the probability of an $n$-sided cell in the whole froth. 
The probability to find a string of length $j$ 
with a termination in a cell with  $n$ sides is consequently given 
by the ratio
\begin{equation}
s_j(n)
= {\sum_{m=3}^\infty N_j(n,m) \over 
\sum_{n,m=3}^\infty N_j(n,m)  } 
=\sum_{m = 3}^\infty C_j(n,m) 
= {N(n) K_j(n) \over N_T \langle K_j \rangle } 
= p(n) { K_j(n) \over \langle K_j \rangle }\;\;\;\;.
\label{sn}
\end{equation}
By following this point of view, the correlation function $C_j(n,m)$ can 
be interpreted as the probability of having a string of 
length $j$ with terminations in two cells with $n$ and $m$ sides respectively.
In an uncorrelated system this conditional probability must be 
the product of the two string probabilities 
\begin{equation}
C^{un}_j(n,m)= s_j(n)s_j(m) =
{ K_j(n) K_j(m) \over \langle K_j \rangle^2 } p(n)p(m) \;\;\;\;.
\label{unCnm}
\end{equation}
Even when the system is uncorrelated, taken two cells 
at distant $j$, the probability to have one cell 
with $m$ sides and the other cell with $n$ sides 
is not given by the simple product of the probabilities of finding 
independently an $n$- and an $m$-sided cells ($p(n)p(m)$).
The factor $K_j(n)K_j(m)/\langle K_j \rangle^2$ in Eq.(\ref{unCnm}) 
indicates that a cell in 
a froth (even in an uncorrelated one) CANNOT be topologically independent by 
its neighbours.
Indeed, in froths, a single isolated cell doesn't exist. 
The cell, its number of sides, and the number of neighbours at any distance 
$j$ constitute a unique system.
(Note that, for $j=1$, the function 
${C_1(n,m) \over C^{un}_1(n,m)}-1=\beta_{n,m}$ is the topological 
short-range order coefficient  introduced by Le Ca\"er et al. \cite{LeCaer}.) 

The structure of a froth can be studied in 
term of effective ``potential'' between different cells. 
These potentials can correspond to a real dynamical 
interactions between coupled cells during the process of 
formation and evolution of the system; or they can 
simply indicate the degrees of affinities of two cells 
to stay at a given relative distance.
Attractive interactions are associated to couples of cells which 
appear in the froth with a higher probability 
than in the uncorrelated case, while negative interactions 
correspond to the opposite case.
Without any loss in generality one can define 
the correlation function of the form
\begin{equation}
C_j(n,m)=C^{un}_j(n,m) \exp \Big(-\beta \varphi_j(n,m) \Big) \;\;\;.
\label{Cjnm2}
\end{equation}
Here $\varphi_j(n,m)$ is the interaction effective potential 
between two cells with $n$ and $m$ sides at topological distance $j$, 
and $\beta$ is the inverse temperature. 
This potential is zero in uncorrelated systems, 
negative when two cells attract and positive when they repel.

We have analyzed data for soap froths prepared at different times, all in the 
scaling regime \cite{Stavans:1,szetotam2}. 
In Fig.1, we show the variation of $C_j(n,m)/C^{un}_j(n,m)$ (Fig.1a) 
and $\varphi_j(n,m)$ (Fig.1b) vs topological distance $j$. 
These correlations and potentials are the same for 
a sample that is 6 hours in the scaling regime as another sample that 
is 14 hours in the scaling regime. 
Other times have also been checked and $C_j(n,m)/C^{un}_j(n,m)$ 
and $\varphi_j(n,m)$ are time independent within experimental errors. 
>From Fig.1a, it is clear that soap froth is strongly correlated at least 
up to the third neighbours. 
Cells with equal number of sides (e.g. $(n,m)=(5,5)$, $(6,6)$, $(7,7)$) 
have lower probability to be first neighbours than 
in the uncorrelated case and the contrary is for cells 
with different number of sides (see Fig.1a for $j=1$).
The associated effective potential between first neighbours is therefore 
repulsive for equal sided cells and attractive 
for cells with different number of sides (see Fig.1b).
This is the Aboav law \cite{Aboav} in the language of correlation.
Increasing the topological distance $j$ the correlation 
and the potential shows oscillations which can be associated 
to the screening and anti-screening of the effective interaction.

In accordance with previous analysis on first neighbours correlation 
\cite{LeCaer,PeshRiv91,DelLeC91}, we introduce the correlator 
$A_j(n,m)\equiv { [ C_j(n,m)/p(n)p(m) ] } $, which is the probability to
have an $n$-sided cell at a distance $j$ from an $m$-sided cell, given that
the two cells exist (this correlator become approximately equal to the
Boltzamnn factor $e^{-\varphi_j(n,m)}$ as $s_j(n) \rightarrow p(n)$ 
for $j\rightarrow \infty$). 
Fig.2 shows the plot of $A_j(n,m)$ vs $m$ for fixed $n=4,..,8, n\ne m$  
and fixed $j=1,2,3$. 
There are theories \cite{rivier,DubRiPe97} that 
predict the bilinearity of $A_1(n,m)$ in $n$, $m$ 
using maximum entropy argument. 
The poor fit to straight line of the data in Fig.2, and the strong deviation
when $n=m$, indicate that the bilinearity of $A_j(n,m)$ is at best a very
crude approximation. 
Afterall, this disagreement should not be too surprising 
as the potential, as defined in Eq.\ref{Cjnm2}, 
manifests strong and long range interaction 
present in soap froth. 
On the other hand, we find a surprisingly simple linear behavior of 
$A_j(n,n)$ vs $n$ for $j=1,2,3$. 
Fig.3 shows the plot of $A_j(n,n)$ vs $n$ for fixed $j=1,2,3$.
The data can be described by the linear function 
$A_j(n,n)=\mu (j) n + \lambda (j)$ with coefficients
$\mu (1)= 0.193 \pm 0.008,\lambda (1)=-0.60\pm 0.05$; 
$\mu (2)= 0.121 \pm 0.005, \lambda (2)= 0.36\pm 0.03$ and  
$\mu (3)= 0.073 \pm 0.003, \lambda (3)= 0.69\pm 0.02$. 

In conclusion, the data we present on soap froth strongly suggest that 
correlation effects are of paramount importance in 
our understanding of the quasi-static and dynamics of froth. 
The data indicate that the correlation is important up to the third shell 
at least,  with  linear correlator for cells of same number of edges, 
but non-bilinear behaviour for cells of different number of edges. 
Future theories of froth dynamics should include these strong correlations. 

\vskip 20pt
\noindent{\bf Acknowledgements}

K.Y. Szeto acknowledges support from the Hong Kong Telecom Institute of 
Information Technology. 
T. Aste acknowledges  discussion with N. Rivier and D. Dubertret and  
the support from the Hong Kong University of Science 
and Technology and the partial support from EU, HCM Program 
``Physics of Foams'' CHRXCT940542 and by the TMR contract ERBFMBICT950380. 
W.Y. Tam acknowledges support from the Direct Allocation Grant 
96/97.SC22 of the Hong Kong University of Science and Technology.

\bigskip

\vskip 20pt
\noindent{\bf Figures}

\noindent{Fig.1 (a) Relative correlation $C_j(n,m)/C^{un}_j(n,m)$ 
and Fig.1 (b) Effective potential $\varphi_j(n,m)$ 
vs topological distance $j$ 
for couples of soap froth cells with $n$ and $m$ sides, and 
$
(n,m)=(5,5), \mbox{\rm solid  circle},\   
(n,m)=(6,5), \mbox{\rm open  square},\  
(n,m)=(6,6), \mbox{\rm solid   triangle},\  
(n,m)=(7,5), \mbox{\rm open  circle},\ 
(n,m)=(7,6), \mbox{\rm open  triangle},\ 
(n,m)=(7,7), \mbox{\rm solid  square}
$}

\noindent{Fig.2 Correlator $A_j(n,m)$ vs $m$ for fixed $n\ne m$,
and (a) $j=1$, (b) $j=2$, and  (c) $j=3$. 
Here $ 
{n=4,\mbox {\rm open  circle},\ } 
{n=5, \mbox{\rm solid  circle},\ }
{n=6, \mbox{\rm open  triangle},\  }
{n=7,\mbox{\rm solid  triangle},\ } 
{n=8,\mbox{\rm open  square}\ }
$
}

\noindent{Fig.3 Correlator $A_j(n,n)$ vs $n$ for fixed $j$;
(a) $j=1, \mbox {\rm open\  circle},\  $
(b) $j=2, \mbox {\rm open\  triangle},\ $
(c) $j=3,\mbox  {\rm open\  square}\ $
}

\newpage
\begin{figure}[!ht]
\vspace{2cm}
\epsfxsize=13.cm
\epsffile{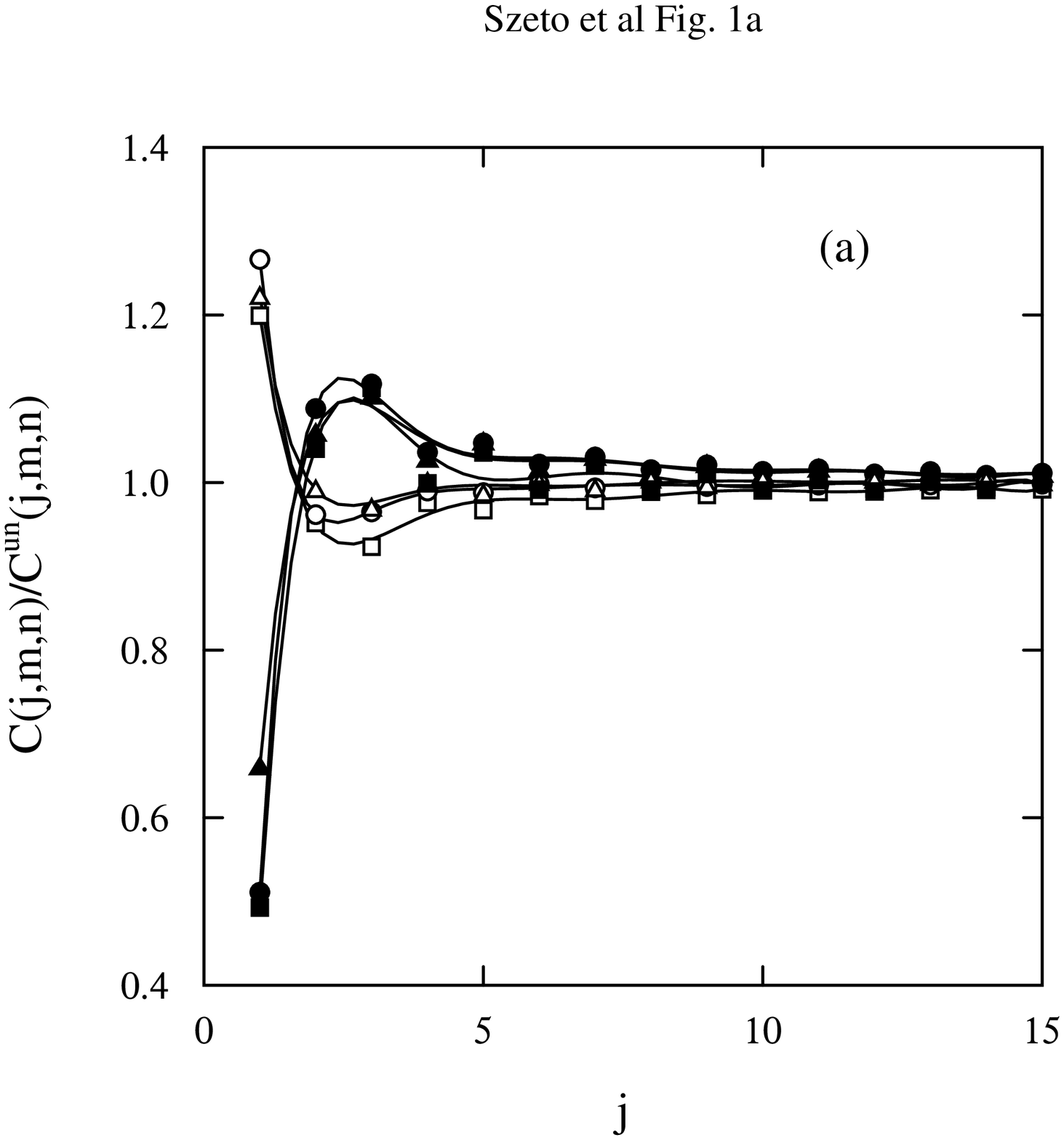}
\begin{centering}
\end{centering}
\end{figure}

\newpage
\begin{figure}[!ht]
\vspace{2cm}
\epsfxsize=13.cm
\epsffile{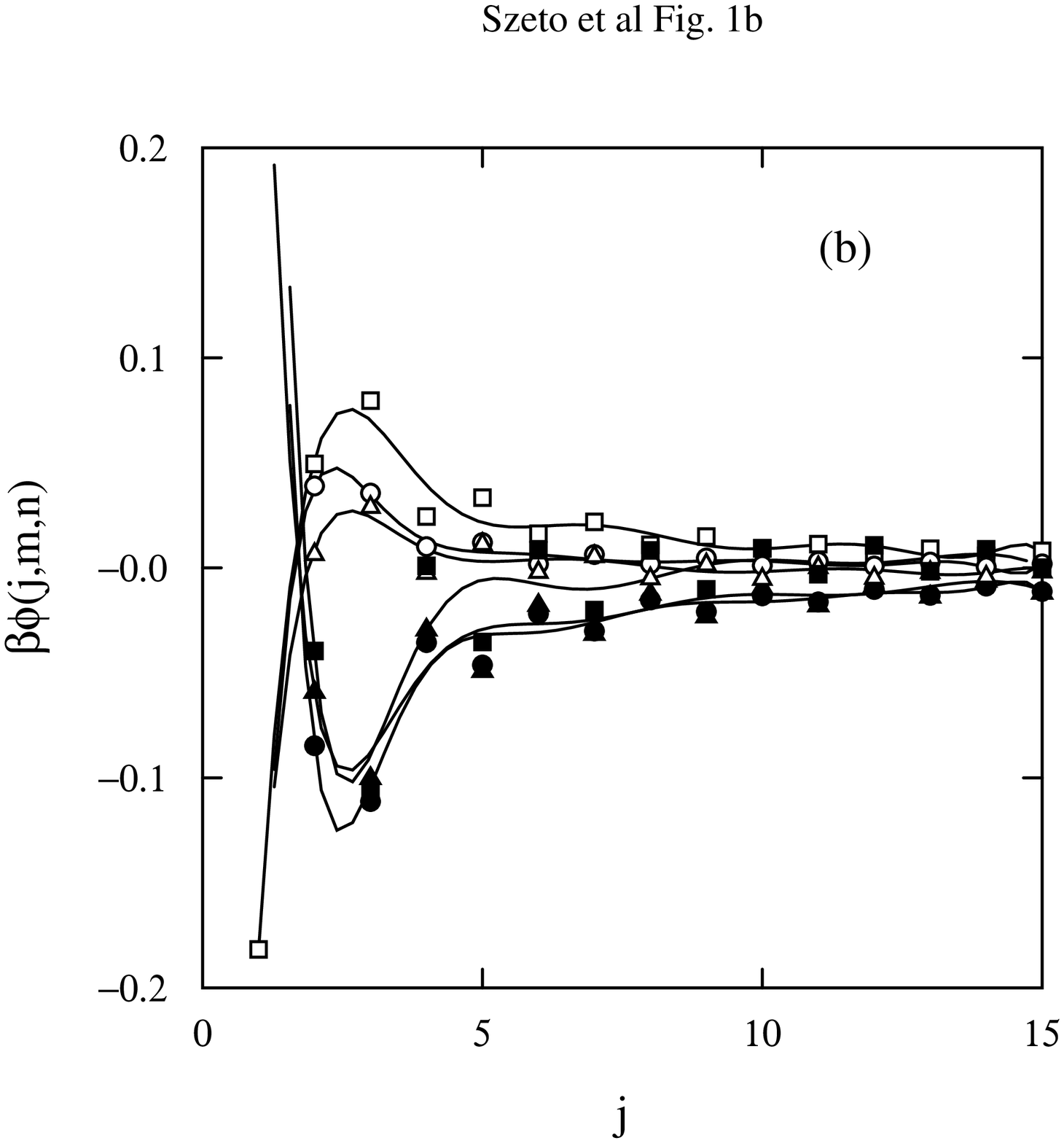}
\begin{centering}
\end{centering}
\end{figure}

\newpage
\begin{figure}[!ht]
\vspace{2cm}
\epsfxsize=13.cm
\epsffile{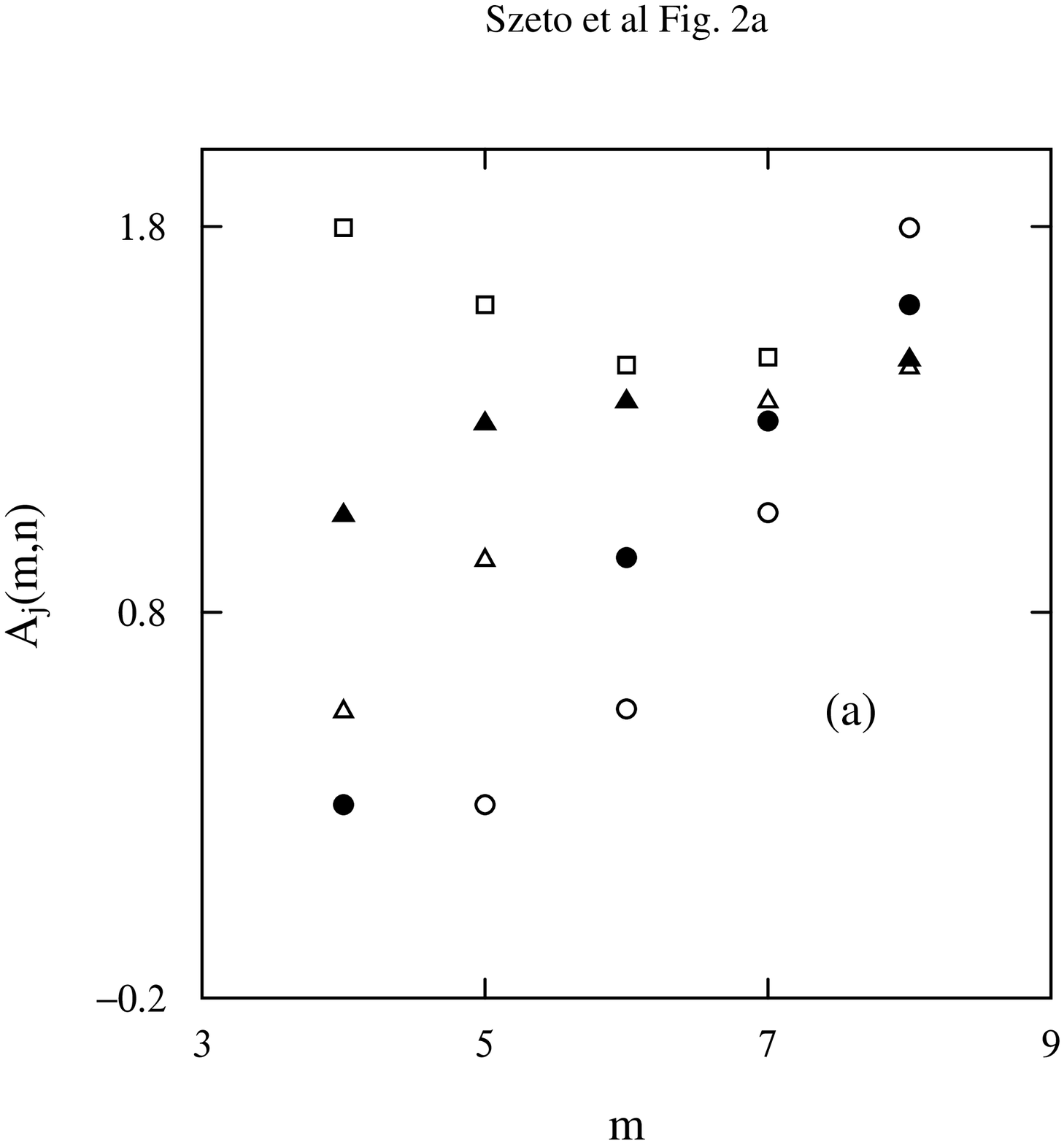}
\begin{centering}
\end{centering}
\end{figure}

\newpage
\begin{figure}[!ht]
\vspace{2cm}
\epsfxsize=13.cm
\epsffile{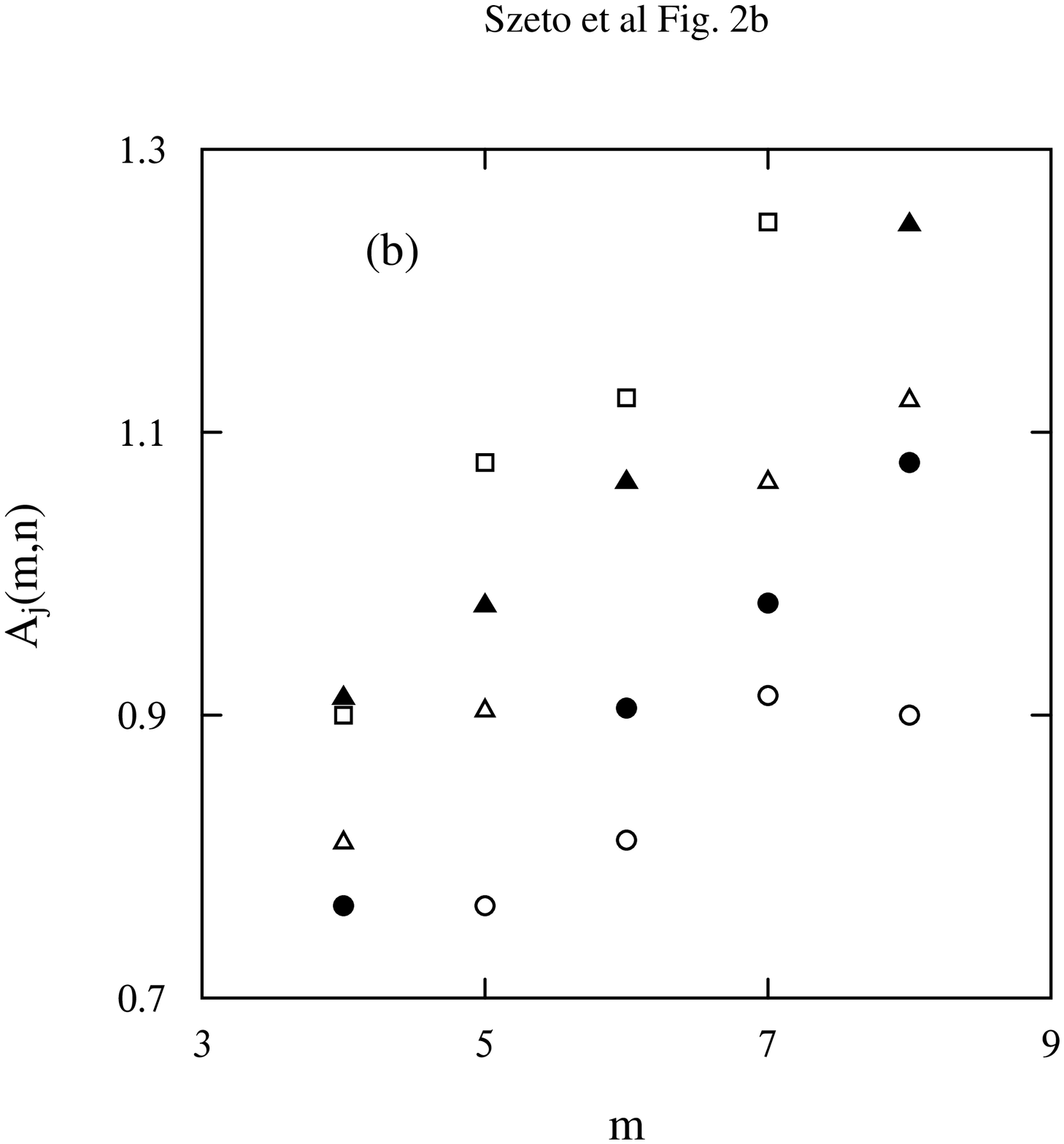}
\begin{centering}
\end{centering}
\end{figure}

\newpage
\begin{figure}[!ht]
\vspace{2cm}
\epsfxsize=13.cm
\epsffile{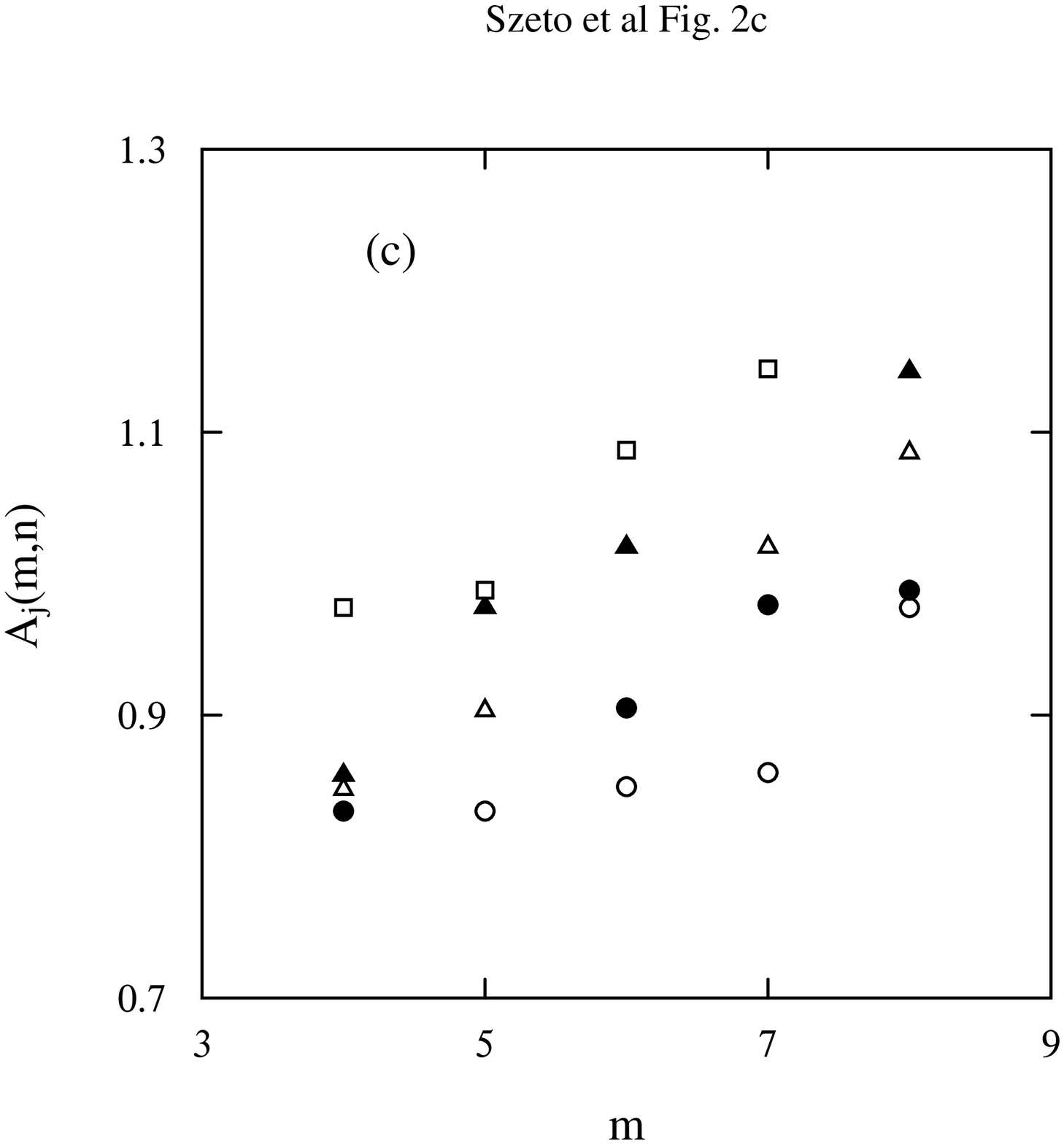}
\begin{centering}
\end{centering}
\end{figure}

\newpage
\begin{figure}[!ht]
\vspace{2cm}
\epsfxsize=13.cm
\epsffile{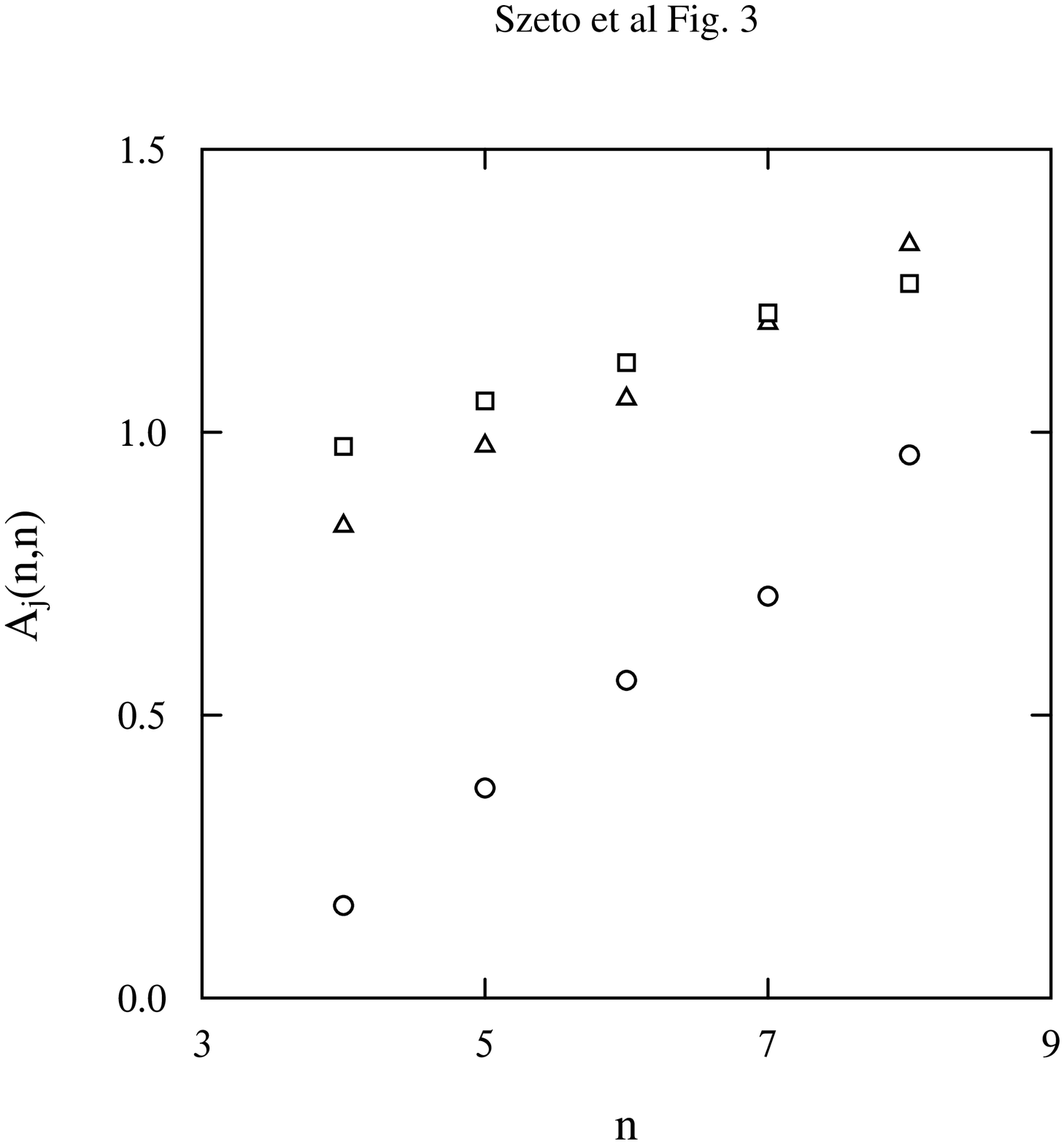}
\begin{centering}
\end{centering}
\end{figure}

\end{document}